\documentclass[aps,prl,reprint]{revtex4-2}
\usepackage{blindtext}
\usepackage{hyperref}
\usepackage[T1]{fontenc}
\usepackage{amsmath}
\usepackage{mathtools}
\usepackage{amsfonts}
\usepackage{amssymb}
\usepackage{graphicx}

\newcommand{\del}{\partial}
\begin{document}
	
	\title{Finite model of an electric charge}
	\author{Suvikranth Gera}
	\email{suvikranthg@iitkgp.ac.in}
	\author{Sandipan Sengupta}
	\email{sandipan@phy.iitkgp.ac.in}
	\affiliation{Department of Physics, Indian Institute of Technology Kharagpur, Kharagpur-721302, India}
\begin{abstract}
We set up a model of an electric charge where the noninvertible metric phase of first order gravity supercedes the point charge singularity in a curved spacetime. A topological interpretation of the electric charge is provided in terms of an index defined for the degenerate spacetime solution, being closely
related to the Euler characteristic. The gravitational equations of motion at this phase are found to be equivalent to the laws of electrostatics. The associated field energy is finite and the geometry sourcing the charge is regular.
\end{abstract}	
\maketitle 
\section{Introduction}
A point particle model for electric charge is inherently singular, and is unsatisfactory for several reasons. In Einstein's general relativity, the point charge leads to divergences both in the field energy and the geometry. Attempts towards alternative constructions that are free of the infinity in the self-energy of the electron have a long history, highlighted by the works of Abraham, Lorentz, Lees, Dirac, Born-Infeld \cite{abraham,*lorentz,*poincare,lees,dirac,born} and so on.  

The present work is concerned with developing a finite model of an elementary charge in curved spacetime. We adopt a geometric perspective, rooted firmly within classical first order gravity theory which can exhibit invertible as well as noninvertible metric phases. 

Firstly, we show that an (apparent) electric charge can be described by a noninvertible metric phase of first order gravity. The solution to the field equations is worked out in detail. The geometry outside this phase is locally equivalent to the Reissner-Nordstr\"{o}m metric upto a coordinate transformation, and hence corresponds to the Einstein phase.

Within this framework, the gravitational equations of motion provide a natural definition for an emergent electric field at the zero-determinant phase. These equations are equivalent to the laws of electrostatics. The associated three-geometry is shown to be regular. We also analyze the field energy in the Hamiltonian form for the full spacetime. In contrast to the point charge model, it is finite. 

Finally, we unravel a topological interpretation of the electric charge in the first order formulation of gravity. It is shown to have a one to one correspondence with a topological invariant defined for degenerate spacetime solutions. This index is integer valued and is closely related to the Euler characteristic, the latter being well-defined only for a non-invertible geometry.

Let us now present our main ideas and results in some detail. 


\section{A charged spacetime solution in Einstein gravity}
In Einsteinian gravity theory, the presence of an electric charge curves the spacetime. For spherical symmetry, the relevant geometry is given by the electrically charged Reissner-Nordstr\"{o}m solution. To begin with, let us consider a closely related metric outside a radius $R > Q$ \cite{gera}:
	\begin{eqnarray}\label{g}
	ds^2&=& -\left[1-\frac{Q}{R(u)}\right]^2 dt^2 + \frac{R^{'2}(u) du^2}{\left[1-\frac{Q}{R(u)}\right]^2}\nonumber\\ 
	&+& R^2(u)(d\theta^2+ \sin ^2\theta d\phi^2)
\end{eqnarray}
where Q is a constant and $R(u)$ is any smooth, monotonically increasing function satisfying the properties below:
 \begin{eqnarray*}
	R(u)\rightarrow Q, R^{(n)}(u)\rightarrow 0 \mathrm{~~as~~}u\rightarrow u_0~(R\rightarrow Q),
\end{eqnarray*}
$R^{(n)}(u)$ being the n-th derivative for any arbitrary integer
$n$. 

For the special parametrization $R(u) = u$, this geometry formally reduces
to the extremal Reissner-Nordstr\"{o}m spacetime at $R > Q$, $Q$ being the (modulus of)electric charge. However, such a coordinate transformation does not satisfy the properties above. In other words, the charged spacetime (\ref{g}) is locally equivalent to the Reissner-Nordstr\"{o}m solution everywhere except at $u=u_0$ ($R=Q$).

The curvature two-forms fields read:
	\begin{eqnarray}
	R^{01}&=&\frac{Q}{R^3}\left(2-\frac{3Q}{R}\right) R'\; dt\wedge du\nonumber\\
	R^{02}&=&-\frac{Q}{R^2}\left(1-\frac{Q}{R}\right)^2 \;dt\wedge d\theta\nonumber\\
	R^{03}&=&-\frac{Q}{R^2}\left(1-\frac{Q}{R}\right)^2\sin\theta \;dt\wedge d\phi \nonumber\\
	R^{12}&=&-\frac{Q}{R^2} R'\;du\wedge d\theta \nonumber\\
	R^{23}&=&\frac{Q}{R}\left(2-\frac{Q}{R}\right)\sin\theta \; d\theta\wedge d\phi\nonumber\\
	R^{31}&=&-\frac{Q}{R^2} R'\sin\theta \; d\phi\wedge du 
\end{eqnarray}
The only component that has a non-trivial limit as $u\rightarrow u_0$ is  $R^{23}\rightarrow \sin\theta \; d\theta\wedge d\phi$.
In terms of the radial coordinate $R$, the Electric  field  is manifestly Coulombic:
\begin{equation}
	E^R\equiv F^{tR}= \frac{Q}{R^2}
\end{equation}

The determinant of the metric above goes to zero smoothly as the surface $u=u_0$ is approached. Thus, the natural continuation that could be defined beyond this surface (at $u<u0$) must involve a noninvertible metric phase \cite{marc,tseytlin,bengtsson,*bengtsson1,sandipan,*sandipan1,kaul}. In the next section, we find the appropriate   noninvertible metric solution of the first order equations of motion in gravity theory which allows such a (smooth) continuation. 

This should be contrasted to the case where the full spacetime exhibits an Einstein phase ($\det g_{\mu\nu}\neq 0$) only. Since the associated spacetime metric there does not degenerate at $R=Q$, 
it could only be continued to an invertible phase of first order gravity. Thus, the metric throughout must correspond to the (electrically charged) Reissner-Nordstr\"{o}m solution of the Einstein-Maxwell theory. In other words, an elementary charge is necessarily associated with a point (curvature) singularity, or equivalently, a divergent (electrostatic) self-energy, at least within the standard formulation of Einstein gravity.

\section{ The non-invertible phase }
This phase in vacuum is defined by any degenerate tetrad solution of the first-order Lagrangian formulation \cite{tseytlin}:
\begin{eqnarray}\label{L}
{\cal L}=\epsilon^{\mu\nu\alpha\beta}\epsilon_{IJKL} e_\mu^I e_\nu^J R_{\alpha\beta}^{~~KL}(\omega)
\end{eqnarray}
The basic field variables are the tetrad and connection, and the indices $\mu$ and $I$ label the spacetime and local Minkowski coordinates, respectively.
Given metric (\ref{g}), a natural choice for its continuation at $R \leq Q~ (u \leq u_0)$ to this phase is the following:
\begin{equation}
	ds^2 = 0+  F^2(u) du^2 + R^2(u)(d\theta^2 +\sin^2\theta d\phi^2)
\end{equation}
The internal metric is $\eta_{IJ}=diag(-1,1,1,1)$. In general, degenerate space-time solutions of first order gravity could exhibit torsion even in the absence of matter fields (e.g. fermions). For our purposes here, it is sufficient to consider the case of vanishing torsion: $T_{\mu\nu}^{~~I} \equiv \frac{1}{2}D_{[\mu}(\omega)e_{\nu]}^{~~I} =0$.

Note that even though the tetrad fields are not invertible, the nontrivial part of the metric above may be used to define a set of triad fields $\hat{e}_a^i\equiv e_a^i$ ($i\equiv[1,2,3],~a\equiv[u,\theta,\phi]$) along with their inverses $\hat{e}^a_i$ ($\hat{e}^a_i \hat{e}_a^j=\delta_i^j,~\hat{e}^b_k \hat{e}_a^k=\delta_a^b$).

The vacuum theory defined by (\ref{L}) corresponds to the following set of field equations:
\begin{eqnarray}\label{eom}
	\epsilon^{\mu\nu\alpha\beta}\epsilon_{IJKL} e_\mu^I T_{\nu\alpha}^{~~J}=0,~ \epsilon^{\mu\nu\alpha\beta}\epsilon_{IJKL} e_\nu^I R_{\alpha\beta}^{~~JK} =0
\end{eqnarray}
For zero torsion, the first set above is satisfied trivially. It is straightforward to see that the  corresponding general solution for the connection fields is given by:
\begin{eqnarray}\label{connectn}
	\omega^{12}&= \left(-\frac{R'}{ F}\right) d\theta,~
	\omega^{23}&=  -\cos\theta d\phi \nonumber\\
	\omega^{31}&= \frac{R'}{F}\sin\theta d\phi ,~ \omega^{0i}&={\cal E}^{ik}e_k \equiv {\cal E}^i
\end{eqnarray}
where ${\cal E}_{kl}={\cal E}_{lk}$ is a symmetric matrix and we have defined $\omega_a^{~0i}\equiv{\cal E}_a^{i}$ in the last equality.
Among the second set of field equations in (\ref{eom}), all components are satisfied identically except $\mu=t,I=0$, which reads:
\begin{eqnarray}\label{mc}
\epsilon^{abc} \epsilon_{ijk}e_a^i R_{bc}^{~~jk}=0&=&\frac{2R}{ F}\left(\frac{R'}{ F}\right)'
+\left(\frac{R'}{ F}\right)^2 \nonumber\\
&-&\frac{1}{2}[{\cal E}_i^{~i}{\cal E}_k^{~k}-{\cal E}^{ik}{\cal E}_{ki}] R^2-1
\end{eqnarray}

Rather than working with the most general solution for ${\cal E}_a^i$ in eq.(\ref{connectn}), we shall use a particular solution in order to be explicit:
\begin{eqnarray}\label{E}
{\cal E}_a^i=\lambda \hat{e}_a^i
\end{eqnarray} 
where $\lambda$ is a constant.
Insertion of this into eq.(\ref{mc}) implies:
\begin{eqnarray*}
\frac{2R}{ F}\left(\frac{R'}{ F}\right)' +\left(\frac{R'}{ F}\right)^2 
-3 \lambda^2 R^2 -1=0
\end{eqnarray*}
This has the following solution:
\begin{equation*}
	F = \frac{R'}{\sqrt{1-\frac{Q}{R}+\lambda^2 R^2}}
\end{equation*}
With this, the solution for the four-metric finally becomes:
\begin{equation}\label{sol}
	ds^2 =0+ \frac{R'^2(u) du^2}{1+\lambda^2 R^2(u)-\frac{Q}{R(u)}}  + R^2(u)(d\theta^2 +\sin^2\theta d\phi^2).
\end{equation}

\section{Electrostatics from degenerate geometry}

The nondegenerate part of the four-metric solution (\ref{sol}) naturally defines an emergent (spatial) three-geometry. This three-metric exhibits an $R\otimes S^2$ topology, having an inner boundary $u=u_*$ ($R=R_*$) defined by the relation $1-\frac{Q}{R_*}+\frac{R_*^2}{Q^2}=0$, where the geometry is nevertheless regular.
Within such an effective description, the geometric (gravitational) fields are completely given by the associated triad $\hat{e}_a^i$ and the (torsionless) spatial connection components $\hat{\omega}_a^{ij}(\hat{e})\equiv\omega_a^{ij}$. 
The remaining components of the spacetime connection $\omega^{0i}$, which are not determined by the emergent triad fields $\hat{e}^i_a$, are not part of this geometry. Hence, these should be interpreted as matter fields. Clearly, these must encode the emergent electric field sourcing the apparent electric charge $Q$ as measured at a large radial distance $R > Q$.

To this end, we consider the equation of motion involving $\omega^{0i}$, given by the $\mu=t,I=i$ component of the second set in eq.(\ref{eom}):
\begin{eqnarray}
\epsilon^{abc} \epsilon_{ijk}e_a^j R_{bc}^{~~0k}=0=\hat{D}_a(\hat{\omega})\left[\epsilon^{abc}\epsilon_{ijk}e_b^j {\cal E}_c^k\right]
\end{eqnarray}
where the three-covariant derivative is defined with respect to the spatial connection $\hat{\omega}_a^{ij}(\hat{e})$. Let us now project this equation along a three-vector $n^i$ in the internal space, leading to:
		 \begin{eqnarray}
	\partial_a\left[\hat{e}({\cal E}_l^{~l} \hat{e}_i^a-{\cal E}_{ik}\hat{e}^{ak})n^i\right]=
	\hat{e}\left[{\cal E}_l^{~l} \hat{e}_i^a-{\cal E}_{ik}\hat{e}^{ak}\right]\hat{D}_a(\hat{\omega})n^i\nonumber\\
	~
\end{eqnarray}
This precisely is equivalent to the Gauss' law $\del_a [\hat{e}E^a]=\hat{e}\rho$ in electrostatics, implying the following definitions for the emergent electric field and the charge density, respectively:
		\begin{eqnarray}\label{emergent}
	E^a&=&\frac{1}{2}( {\cal E}_l^{~l} \hat{e}_i^a-{\cal E}_{ik}\hat{e}^{ak})n^i=\frac{1}{2}n^{[a}\hat{e}^{b]}_k \omega_b^{~0k},\nonumber\\
	\rho&=&\frac{1}{2}({\cal E}_l^{~l} \hat{e}_i^a-{\cal E}_{ik}\hat{e}^{ak})\hat{D}_a(\hat{\omega})n^i
\end{eqnarray}
where we have defined: $n^a\equiv n^i \hat{e}_i^a$.
Note that the expression for the emergent U(1) electric field is gauge-invariant. Also, it satisfies $\epsilon^{abc}\del_a E_b=0$ trivially, as it should.

Let us now apply the general construction above to the special case where the field is non-trivial only along the radial direction. This may be achieved by choosing the vector $n^i$ to be normal to the internal two-sphere; $n^i\equiv (1, 0, 0)$. 
Using eq.(\ref{E}) and transforming to the radial coordinate $R$ inside the phase boundary (at $u<u_0$), the resulting emergent field then reads:
\begin{equation}
E^R=\lambda \sqrt{1-\frac{Q}{R}+\lambda^2 R^2},~E^\theta=0=E^\phi
\end{equation}
Continuity of the field across the phase boundary $u = u_0$ fixes the dimensionful constant as $\lambda=\frac{1}{Q}$. 
The static field $E^R(R)$ at $R\leq Q$ is not Coulombic, even though the field outside the phase (outer) boundary is. It vanishes at the inner boundary $R = R_*$. This is a crucial difference compared to the point charge model. 

In the limit $u\rightarrow u_0$ as one approaches the phase boundary, the electric flux through the two sphere reads:
\begin{eqnarray}
	\frac{1}{4\pi}\int_{S^2}d^2 x~ E^a n_a  =\frac{1}{4\pi Q}\int_{S^2} d\theta d\phi \sin\theta  R^2= Q,
\end{eqnarray}
which is the same as the apparent electric charge defining the invertible phase at $R>Q$.

\section{Finiteness of field-energy}
Let us consider the Hamiltonian form of the electrostatic self-energy for an electrically charged configuration in a fixed curved spacetime. For the invertible phase, the relevant part of the Lagrangian density is the Maxwell term: ${\cal L}_m=-\frac{1}{4}eF^{\alpha\beta}F_{\alpha\beta}$. We shall use the standard reparametrization of the metric (\ref{g}) in terms of lapse ($N$) and shift ($N^a$) variables, the latter being trivial in this case:
\begin{eqnarray}
e_t^I=NM^I,~e_a^I M_I=0,~q_{ab}=e_a^I e_{bI},~e=N\sqrt{q}
\end{eqnarray}
Using the expression for the canonical momenta $\pi^a=-eF^{ta}=N\sqrt{q}E^a$ in terms of the electric field $E^a$, the Hamiltonian for a vanishing magnetic field ($F_{ab}=0$) reads
\begin{eqnarray}\label{H}
{\cal H}=\frac{N^3\sqrt{q}}{2}q_{ab}E^a E^b
\end{eqnarray}
In obtaining this expression, the Gauss' law has already been implemented as a constraint.
The electrostatic energy associated with the invertible phase at $Q<R<\infty$ is then equal to the Hamiltonian $H$ upto a normalization:
\begin{eqnarray}
H&=&\int d^3 x ~{\cal H}=\frac{1}{2}\int dR ~d\theta~ d\phi~ \sin\theta R^2 (E^R)^2\nonumber\\
&=&2\pi \left[\frac{Q^2}{R}\right]_{\infty}^{Q}
=2\pi Q
\end{eqnarray}

The non-invertible phase at $R\leq Q$, on the other hand, has no electrostatic energy in the sense above. This fact is reflected by the fact that the Hamiltonian density (\ref{H}) vanishes as the lapse $N$ goes to zero (smoothly) with $q_{ab}$ and $E^a$ being finite, as is the case here. Thus, the degenerate phase with a vanishing lapse does not contribute to the self-energy in this limiting sense. 

To emphasize, our geometric model of an electric charge exhibits a finite self-energy at the classical level. This is in contrast to standard electrostatics where an infinite amount of energy is needed to assemble a finite charge within a point. This divergence is precisely what gets reflected in the Hamiltonian evaluated for a (extremal) Reissner-Nordstr\"{o}m geometry defining the whole spacetime $0<R<\infty$: $H=2\pi \left[\frac{Q^2}{R}\right]_{\infty}^{0}\rightarrow \infty$.

\section{Regularity of geometry}
At this point a relevant question is, whether the emergent three-geometry above could be singular, even though the field energy itself is manifestly not. An explicit evaluation of all the three-curvature scalars, however, shows that the geometry is in fact regular. For instance, let us display the scalars upto quadratic order at $R\leq Q$ explicitly,
	\begin{eqnarray}
		&&{\cal R}^{~~ab}_{ab}=\frac{1}{R^2}\left(\frac{-2R}{ F}\left(\frac{R'}{ F}\right)'-\left(\frac{R'}{ F}\right)^2+3\lambda^2 R^2 +1\right)\nonumber\\
		&&~=0 ;\nonumber\\
		&&{\cal R}^{ab}{\cal R}_{ab}= \frac{3Q^2}{2R^6}
	\end{eqnarray}
 
Finally, since torsion is trivial, all the corresponding three-scalars  vanish:
\begin{equation*}
	\epsilon^{abc}T_{abc}=0=T_{abc}T^{abc}.
\end{equation*}
Hence, there is no curvature or torsion singularity in the three-geometry.

\section{Topological interpretation of electric charge}
In four dimensions, there exist three topological invariants in first order gravity theory. For vanishing spatial contortion $K_a^{~ij}$ at the degenerate core, the parity odd topological invariants, given by Nieh-Yan and Pontryagin numbers, are both trivial. The only other  invariant is the Euler index (parity even), which is not well-defined for a degenerate spacetime. 

Here we define a topological index to characterize  the degenerate geometry relevant to the context here. This is necessarily nontrivial for a nonvanishing electric charge at the Einstein (Reissner-Nordstr\"{o}m) phase. With this, the
electric charge acquires a novel topological interpretation in gravity theory.

From the field-strength components associated with the noninvertible phase, the Euler density is found to vanish (since $R^{0i}=\lambda D(\omega) e^i=0$):
\begin{eqnarray}\label{eu}
I_E = \frac{1}{32\pi^2} \epsilon_{IJKL}R^{IJ}\wedge R^{KL} =0.
\end{eqnarray}
 This, however, must be considered along with the possible boundary corrections, for which a suitable prescription is required. For the degenerate solution discussed earlier, $g_{tu} = 0$ at $R\leq Q$ and the time coordinate $t$ ceases to evolve. This implies that we could consider a spacetime which allows a smooth limit $g_{tt}\rightarrow 0$ and is associated with a trivial (bulk) Euler density as above. It is then possible  to define the boundary correction to Euler invariant based on this spacetime, before taking the degenerate limit $g_{tt}\rightarrow 0$ at the end. The resulting geometric (topological) invariant, if nontrivial, could be seen as a genuine characteristic of the degenerate spacetime. 

To this end, we define the following spacetime metric which reduces to the degenerate geometry (\ref{sol}) as $\epsilon(t\rightarrow t_*)\rightarrow 0$ for some $t_*$:
\begin{equation*}
	ds^2 = -\epsilon^2(t) dt^2 +\frac{dR^2}{1-\frac{Q}{R}+\frac{R^2}{Q^2}}+ R^2(d\theta^2 +\sin^2 \theta d\phi^2)
\end{equation*}
The associated connection fields (torsionless) read:
\begin{equation*}
	\bar{\omega}^{ij}=\omega^{ij},~ \bar{\omega}^{0i}=0.
\end{equation*}
where $\omega^{ij}$ are given by eq.(\ref{connectn}). Note that the associated Euler density vanishes exactly as in (\ref{eu}), since $\bar{R}^{0i}(\bar{\omega})=0$.

For evaluating the contribution of the $t=t_*$ boundary to the Euler number, a relevant variable is $\theta^{IJ}= \bar{\omega}^{IJ}-\omega^{IJ}$ \cite{eguchi,*gibbons}:
\begin{equation*}
	\theta^{ij}=0,~ \theta^{0i}=\lambda e^i
\end{equation*}
Using the appropriate normalization associated with the radial direction, the boundary correction (in analogy to the Euler boundary correction) may be defined  as:
\begin{eqnarray}
&\chi_B& \nonumber\\
&=&-\frac{1}{4 V_R}\int_{\partial V}\epsilon_{IJKL}\left[\theta^{IJ}\wedge \bar{R}^{KL}-\frac{2}{3}\theta^{IJ}\wedge	\theta^{K}_{~M}\wedge
	\theta^{ML}\right]\nonumber\\
	&=& \frac{2}{Q^3 V_R}\int_{0}^{2\pi} d\phi \int_0^{\pi} d\theta sin\theta \int_{R_*}^Q dR~\frac{R^2}{\sqrt{1-\frac{Q}{R}+\frac{R^2}{Q^2}}}\nonumber\\
	& =&2
\end{eqnarray}
where $V_R =\frac{4\pi}{Q^3}\int\frac{R^2 dR}{\sqrt{1-\frac{Q}{R}+\frac{R^2}{Q^2}}}$ is the (dimensionless) proper volume of the three-manifold in question. Importantly, this result is independent of $\epsilon$, which may now be taken to be zero. Finally, the sum of the volume and boundary contributions to $\chi$ leads to:
\begin{equation}
	\chi = \chi_V + \chi_B =0+2=2. 
\end{equation} 

It is interesting to note that this index is the same as the Euler characteristic of the compact even-dimensional subspace without boundary, which happens to be a two-sphere in this special case.

Importantly, the emergent electric field vanishes for $\lambda = 0$, which implies a trivial $\chi$-number. Thus, a nonvanishing (apparent) charge $Q$ at the Einsteinian phase ($R>Q$) essentially has a topological origin in the degenerate phase ($R\leq Q$). 

\section{CONCLUDING REMARKS}
We have set up an alternative to the point model of an elementary electric charge. Such a configuration is described by a noninvertible solution of first order field equations in gravity theory. The emergent  electrostatic energy as well as the geometry is finite, superceding the generic divergences of a pointlike model. 

We demonstrate that a nontrivial electric charge in the Einsteinian phase is in fact equivalent to a nonvanishing value for a geometric invariant $\chi$ analogous to the Euler index. These two phases together, alongwith the associated solutions to the first order field equations, define the full spacetime which is smooth everywhere. Thus, the electric charge acquires a topological meaning in gravity theory in absence of matter.  Thus, the electric charge acquires a topological meaning in gravity theory. This configuration has vanishing torsion and does not involve any magnetic charge or current \cite{gera,gera1}. It could be interesting to investigate a possible realization of dyons or a (topological) quantization of charge within this formulation, something that is beyond the scope of the present work. 

It should be emphasized that the framework presented here is different from  instances where charge could be an artefact of multiply connected geometries such as geons or
wormholes \cite{geon1,geon2,sorkin}. In general, these do not thrive on a degenerate tetrad phase.

To conclude, these new features mentioned above, alongwith the manifest role of degenerate spacetime solutions as a possible regulator of divergences, are intriguing enough as they are. It seems to be an open question whether a generic connection between first order gravity and elementary particles could be envisaged along these lines, potentially leading to pathways beyond the standard edifices built upon quantum fields.

\section{Acknowledgments:}
The work of S. S. is partially supported by the ISIRD project Grant (RAQ). S. G. gratefully acknowledges
the support of a DST Inspire doctoral fellowship.

	\bibliographystyle{apsrev4-2}
	\bibliography{reff}

\begin{thebibliography}{20}%
\makeatletter
\providecommand \@ifxundefined [1]{%
 \@ifx{#1\undefined}
}%
\providecommand \@ifnum [1]{%
 \ifnum #1\expandafter \@firstoftwo
 \else \expandafter \@secondoftwo
 \fi
}%
\providecommand \@ifx [1]{%
 \ifx #1\expandafter \@firstoftwo
 \else \expandafter \@secondoftwo
 \fi
}%
\providecommand \natexlab [1]{#1}%
\providecommand \enquote  [1]{``#1''}%
\providecommand \bibnamefont  [1]{#1}%
\providecommand \bibfnamefont [1]{#1}%
\providecommand \citenamefont [1]{#1}%
\providecommand \href@noop [0]{\@secondoftwo}%
\providecommand \href [0]{\begingroup \@sanitize@url \@href}%
\providecommand \@href[1]{\@@startlink{#1}\@@href}%
\providecommand \@@href[1]{\endgroup#1\@@endlink}%
\providecommand \@sanitize@url [0]{\catcode `\\12\catcode `\$12\catcode
  `\&12\catcode `\#12\catcode `\^12\catcode `\_12\catcode `\%12\relax}%
\providecommand \@@startlink[1]{}%
\providecommand \@@endlink[0]{}%
\providecommand \url  [0]{\begingroup\@sanitize@url \@url }%
\providecommand \@url [1]{\endgroup\@href {#1}{\urlprefix }}%
\providecommand \urlprefix  [0]{URL }%
\providecommand \Eprint [0]{\href }%
\providecommand \doibase [0]{https://doi.org/}%
\providecommand \selectlanguage [0]{\@gobble}%
\providecommand \bibinfo  [0]{\@secondoftwo}%
\providecommand \bibfield  [0]{\@secondoftwo}%
\providecommand \translation [1]{[#1]}%
\providecommand \BibitemOpen [0]{}%
\providecommand \bibitemStop [0]{}%
\providecommand \bibitemNoStop [0]{.\EOS\space}%
\providecommand \EOS [0]{\spacefactor3000\relax}%
\providecommand \BibitemShut  [1]{\csname bibitem#1\endcsname}%
\let\auto@bib@innerbib\@empty
\bibitem [{\citenamefont {Abraham}(1902)}]{abraham}%
  \BibitemOpen
  \bibfield  {author} {\bibinfo {author} {\bibfnamefont {M.}~\bibnamefont
  {Abraham}},\ }\href@noop {} {\bibfield  {journal} {\bibinfo  {journal} {Phys.
  Zeit}\ }\textbf {\bibinfo {volume} {4}},\ \bibinfo {pages} {57} (\bibinfo
  {year} {1902})}\BibitemShut {NoStop}%
\bibitem [{\citenamefont {Lorentz}(1952)}]{lorentz}%
  \BibitemOpen
  \bibfield  {author} {\bibinfo {author} {\bibfnamefont {H.}~\bibnamefont
  {Lorentz}},\ }\href@noop {} {\bibinfo {title} {Theory of electrons, 2nd
  edn.(1915)}} (\bibinfo {year} {1952})\BibitemShut {NoStop}%
\bibitem [{\citenamefont {Poincar{\'e}}(1905)}]{poincare}%
  \BibitemOpen
  \bibfield  {author} {\bibinfo {author} {\bibfnamefont {H.}~\bibnamefont
  {Poincar{\'e}}},\ }\href@noop {} {\bibfield  {journal} {\bibinfo  {journal}
  {CR de l’Ac. des Sci. de Paris}\ }\textbf {\bibinfo {volume} {40}},\
  \bibinfo {pages} {1504} (\bibinfo {year} {1905})}\BibitemShut {NoStop}%
\bibitem [{\citenamefont {Ph.D.}(1939)}]{lees}%
  \BibitemOpen
  \bibfield  {author} {\bibinfo {author} {\bibfnamefont {A.~L.~M.}\
  \bibnamefont {Ph.D.}},\ }\href {https://doi.org/10.1080/14786443908521195}
  {\bibfield  {journal} {\bibinfo  {journal} {The Lond, Edinburgh, and Dublin
  PMJS}\ }\textbf {\bibinfo {volume} {28}},\ \bibinfo {pages} {385} (\bibinfo
  {year} {1939})}\BibitemShut {NoStop}%
\bibitem [{\citenamefont {Dirac}(1962)}]{dirac}%
  \BibitemOpen
  \bibfield  {author} {\bibinfo {author} {\bibfnamefont {P.~A.~M.}\
  \bibnamefont {Dirac}},\ }\href {https://doi.org/10.1098/rspa.1962.0124}
  {\bibfield  {journal} {\bibinfo  {journal} {Proc of the Royal Society}\
  }\textbf {\bibinfo {volume} {268}},\ \bibinfo {pages} {57} (\bibinfo {year}
  {1962})}\BibitemShut {NoStop}%
\bibitem [{\citenamefont {Born}\ \emph {et~al.}(1934)\citenamefont {Born},
  \citenamefont {Infeld},\ and\ \citenamefont {Fowler}}]{born}%
  \BibitemOpen
  \bibfield  {author} {\bibinfo {author} {\bibfnamefont {M.}~\bibnamefont
  {Born}}, \bibinfo {author} {\bibfnamefont {L.}~\bibnamefont {Infeld}},\ and\
  \bibinfo {author} {\bibfnamefont {R.~H.}\ \bibnamefont {Fowler}},\ }\href
  {https://doi.org/10.1098/rspa.1934.0059} {\bibfield  {journal} {\bibinfo
  {journal} {Proc of the Royal Society}\ }\textbf {\bibinfo {volume} {144}},\
  \bibinfo {pages} {425} (\bibinfo {year} {1934})}\BibitemShut {NoStop}%
\bibitem [{\citenamefont {Gera}\ and\ \citenamefont {Sengupta}(2020)}]{gera}%
  \BibitemOpen
  \bibfield  {author} {\bibinfo {author} {\bibfnamefont {S.}~\bibnamefont
  {Gera}}\ and\ \bibinfo {author} {\bibfnamefont {S.}~\bibnamefont
  {Sengupta}},\ }\href@noop {} {\bibinfo {title} {Emergent monopoles and
  magnetic charge}} (\bibinfo {year} {2020}),\ \Eprint
  {https://arxiv.org/abs/2004.13083} {arXiv:2004.13083 [gr-qc]} \BibitemShut
  {NoStop}%
\bibitem [{\citenamefont {Henneaux}(1979)}]{marc}%
  \BibitemOpen
  \bibfield  {author} {\bibinfo {author} {\bibfnamefont {M.}~\bibnamefont
  {Henneaux}},\ }\href@noop {} {\bibfield  {journal} {\bibinfo  {journal}
  {Bull. Soc. Math. Belg}\ }\textbf {\bibinfo {volume} {31}},\ \bibinfo {pages}
  {47} (\bibinfo {year} {1979})}\BibitemShut {NoStop}%
\bibitem [{\citenamefont {Tseytlin}(1982)}]{tseytlin}%
  \BibitemOpen
  \bibfield  {author} {\bibinfo {author} {\bibfnamefont {A.~A.}\ \bibnamefont
  {Tseytlin}},\ }\href {https://doi.org/10.1088/0305-4470/15/3/005} {\bibfield
  {journal} {\bibinfo  {journal} {Journal of Physics A: Mathematical and
  General}\ }\textbf {\bibinfo {volume} {15}},\ \bibinfo {pages} {L105}
  (\bibinfo {year} {1982})}\BibitemShut {NoStop}%
\bibitem [{\citenamefont {BENGTSSON}(1989)}]{bengtsson}%
  \BibitemOpen
  \bibfield  {author} {\bibinfo {author} {\bibfnamefont {I.}~\bibnamefont
  {BENGTSSON}},\ }\href {https://doi.org/10.1142/S0217751X89002363} {\bibfield
  {journal} {\bibinfo  {journal} {International Journal of Modern Physics A}\
  }\textbf {\bibinfo {volume} {04}},\ \bibinfo {pages} {5527} (\bibinfo {year}
  {1989})},\ \Eprint
  {https://arxiv.org/abs/https://doi.org/10.1142/S0217751X89002363}
  {https://doi.org/10.1142/S0217751X89002363} \BibitemShut {NoStop}%
\bibitem [{\citenamefont {Bengtsson}(1991)}]{bengtsson1}%
  \BibitemOpen
  \bibfield  {author} {\bibinfo {author} {\bibfnamefont {I.}~\bibnamefont
  {Bengtsson}},\ }\href {https://doi.org/10.1088/0264-9381/8/10/010} {\bibfield
   {journal} {\bibinfo  {journal} {Classical and Quantum Gravity}\ }\textbf
  {\bibinfo {volume} {8}},\ \bibinfo {pages} {1847} (\bibinfo {year}
  {1991})}\BibitemShut {NoStop}%
\bibitem [{\citenamefont {Kaul}\ and\ \citenamefont
  {Sengupta}(2016{\natexlab{a}})}]{sandipan}%
  \BibitemOpen
  \bibfield  {author} {\bibinfo {author} {\bibfnamefont {R.~K.}\ \bibnamefont
  {Kaul}}\ and\ \bibinfo {author} {\bibfnamefont {S.}~\bibnamefont
  {Sengupta}},\ }\href {https://doi.org/10.1103/PhysRevD.93.084026} {\bibfield
  {journal} {\bibinfo  {journal} {Phys. Rev. D}\ }\textbf {\bibinfo {volume}
  {93}},\ \bibinfo {pages} {084026} (\bibinfo {year}
  {2016}{\natexlab{a}})}\BibitemShut {NoStop}%
\bibitem [{\citenamefont {Kaul}\ and\ \citenamefont
  {Sengupta}(2016{\natexlab{b}})}]{sandipan1}%
  \BibitemOpen
  \bibfield  {author} {\bibinfo {author} {\bibfnamefont {R.~K.}\ \bibnamefont
  {Kaul}}\ and\ \bibinfo {author} {\bibfnamefont {S.}~\bibnamefont
  {Sengupta}},\ }\href {https://doi.org/10.1103/PhysRevD.94.104047} {\bibfield
  {journal} {\bibinfo  {journal} {Phys. Rev. D}\ }\textbf {\bibinfo {volume}
  {94}},\ \bibinfo {pages} {104047} (\bibinfo {year}
  {2016}{\natexlab{b}})}\BibitemShut {NoStop}%
\bibitem [{\citenamefont {Kaul}\ and\ \citenamefont {Sengupta}(2017)}]{kaul}%
  \BibitemOpen
  \bibfield  {author} {\bibinfo {author} {\bibfnamefont {R.}~\bibnamefont
  {Kaul}}\ and\ \bibinfo {author} {\bibfnamefont {S.}~\bibnamefont
  {Sengupta}},\ }\href {https://doi.org/10.1103/PhysRevD.96.104011} {\bibfield
  {journal} {\bibinfo  {journal} {Phys. Rev. D}\ }\textbf {\bibinfo {volume}
  {96}},\ \bibinfo {pages} {104011} (\bibinfo {year} {2017})}\BibitemShut
  {NoStop}%
\bibitem [{\citenamefont {Eguchi}\ \emph {et~al.}(1980)\citenamefont {Eguchi},
  \citenamefont {Gilkey},\ and\ \citenamefont {Hanson}}]{eguchi}%
  \BibitemOpen
  \bibfield  {author} {\bibinfo {author} {\bibfnamefont {T.}~\bibnamefont
  {Eguchi}}, \bibinfo {author} {\bibfnamefont {P.~B.}\ \bibnamefont {Gilkey}},\
  and\ \bibinfo {author} {\bibfnamefont {A.~J.}\ \bibnamefont {Hanson}},\
  }\href {https://doi.org/https://doi.org/10.1016/0370-1573(80)90130-1}
  {\bibfield  {journal} {\bibinfo  {journal} {Physics Reports}\ }\textbf
  {\bibinfo {volume} {66}},\ \bibinfo {pages} {213 } (\bibinfo {year}
  {1980})}\BibitemShut {NoStop}%
\bibitem [{\citenamefont {Gibbons}\ and\ \citenamefont
  {Kallosh}(1995)}]{gibbons}%
  \BibitemOpen
  \bibfield  {author} {\bibinfo {author} {\bibfnamefont {G.~W.}\ \bibnamefont
  {Gibbons}}\ and\ \bibinfo {author} {\bibfnamefont {R.~E.}\ \bibnamefont
  {Kallosh}},\ }\href {https://doi.org/10.1103/PhysRevD.51.2839} {\bibfield
  {journal} {\bibinfo  {journal} {Phys. Rev. D}\ }\textbf {\bibinfo {volume}
  {51}},\ \bibinfo {pages} {2839} (\bibinfo {year} {1995})}\BibitemShut
  {NoStop}%
\bibitem [{\citenamefont {Gera}\ and\ \citenamefont {Sengupta}(2019)}]{gera1}%
  \BibitemOpen
  \bibfield  {author} {\bibinfo {author} {\bibfnamefont {S.}~\bibnamefont
  {Gera}}\ and\ \bibinfo {author} {\bibfnamefont {S.}~\bibnamefont
  {Sengupta}},\ }\href {https://doi.org/10.1103/PhysRevD.99.124038} {\bibfield
  {journal} {\bibinfo  {journal} {Phys. Rev. D}\ }\textbf {\bibinfo {volume}
  {99}},\ \bibinfo {pages} {124038} (\bibinfo {year} {2019})}\BibitemShut
  {NoStop}%
\bibitem [{\citenamefont {Misner}\ and\ \citenamefont {Wheeler}(1957)}]{geon1}%
  \BibitemOpen
  \bibfield  {author} {\bibinfo {author} {\bibfnamefont {C.~W.}\ \bibnamefont
  {Misner}}\ and\ \bibinfo {author} {\bibfnamefont {J.~A.}\ \bibnamefont
  {Wheeler}},\ }\href
  {https://doi.org/https://doi.org/10.1016/0003-4916(57)90049-0} {\bibfield
  {journal} {\bibinfo  {journal} {Annals of Physics}\ }\textbf {\bibinfo
  {volume} {2}},\ \bibinfo {pages} {525 } (\bibinfo {year} {1957})}\BibitemShut
  {NoStop}%
\bibitem [{\citenamefont {Wheeler}(1957)}]{geon2}%
  \BibitemOpen
  \bibfield  {author} {\bibinfo {author} {\bibfnamefont {J.~A.}\ \bibnamefont
  {Wheeler}},\ }\href
  {https://doi.org/https://doi.org/10.1016/0003-4916(57)90050-7} {\bibfield
  {journal} {\bibinfo  {journal} {Annals of Physics}\ }\textbf {\bibinfo
  {volume} {2}},\ \bibinfo {pages} {604 } (\bibinfo {year} {1957})}\BibitemShut
  {NoStop}%
\bibitem [{\citenamefont {Sorkin}(1977)}]{sorkin}%
  \BibitemOpen
  \bibfield  {author} {\bibinfo {author} {\bibfnamefont {R.}~\bibnamefont
  {Sorkin}},\ }\href {https://doi.org/10.1088/0305-4470/10/5/006} {\bibfield
  {journal} {\bibinfo  {journal} {Journal of Physics A: Mathematical and
  General}\ }\textbf {\bibinfo {volume} {10}},\ \bibinfo {pages} {717}
  (\bibinfo {year} {1977})}\BibitemShut {NoStop}%
\end{thebibliography}%

\end{document}